\documentclass[prl,twocolumn,amsmath,floatfix,showpacs]{revtex4}
\usepackage{graphicx}
\usepackage{dcolumn}

\begin{document}
\title{Vibrationally Induced Two-Level Systems in Single-Molecule Junctions}

\author{W.H.A. Thijssen}
\affiliation{Kamerlingh Onnes Laboratorium, Universiteit Leiden,
Niels Bohrweg 2, 2333 CA Leiden, Netherlands}

\author{D. Djukic}
\affiliation{Kamerlingh Onnes Laboratorium, Universiteit Leiden,
Niels Bohrweg 2, 2333 CA Leiden, Netherlands}

\author{A.F. Otte}
\affiliation{Kamerlingh Onnes Laboratorium, Universiteit Leiden,
Niels Bohrweg 2, 2333 CA Leiden, Netherlands}

\author{R.H. Bremmer}
\affiliation{Kamerlingh Onnes Laboratorium, Universiteit Leiden,
Niels Bohrweg 2, 2333 CA Leiden, Netherlands}

\author{J.M. van Ruitenbeek}
\affiliation{Kamerlingh Onnes Laboratorium, Universiteit Leiden,
Niels Bohrweg 2, 2333 CA Leiden, Netherlands}

\begin{abstract}
Single-molecule junctions are found to show anomalous spikes in
dI/dV spectra. The position in energy of the spikes are related to
local vibration mode energies. A model of vibrationally induced
two-level systems reproduces the data very well. This mechanism is
expected to be quite general for single-molecule junctions. It
acts as an intrinsic amplification mechanism for local vibration
mode features and may be exploited as a new spectroscopic tool.
\end{abstract}

\date{\today}
 \maketitle

A single atom or molecule with an almost transparent single
conductance channel leading to a conductance near the conductance
quantum  2e$^2$/h (= 1 G$_{0}$) can be contacted to leads.
Conduction electrons can pass through such junction ballistically
for low bias voltages since the mean free path of the electrons is
much larger than the size of the contact. However the contact is
not entirely ballistic in the sense that once the excess energy of
the conduction electrons becomes equal or larger than the energy
of a local mode of the contact, the electrons can scatter
inelastically by exciting a local mode. This results in the case
of a perfectly transmitting single channel contact to a small
decrease in the conductance, since the forward travelling
electrons are backscattered due to the energy loss in the
inelastic scattering process. Differential conductance (dI/dV)
measurements have identified vibration modes of single molecules
in an atomic contact \cite{smit02}. This technique, also called
Point Contact Spectroscopy (PCS) is analogous to inelastic
electron tunnelling spectroscopy (IETS) for single molecules
\cite{stipe98,park00}, with the difference that the conductance in
the latter case increases due to the opening of an additional
conductance channel.

In this letter we present the observation of anomalous spikes,
rather than steps, in dI/dV measurements on various single
molecule contacts. We present a model that involves two-level
systems, which describes our data very well. It may be used as a
new spectroscopic tool for identifying molecular vibration modes
in single molecule junctions.

\begin{figure} [b!]
\includegraphics[width=8.7cm,angle=0] {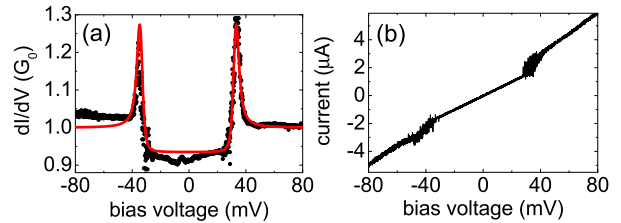}
\caption{\label{fig1} (Color online) (a) dI/dV spectrum on a CO
molecule bridging a Pt contact displaying symmetric positive
spikes. The red curve shows a fit by the model discussed in the
text \cite{fits}; (b) I-V measurement on a CO molecule bridging a
Pt contact}
\end{figure}

We create atomic contacts using a mechanically controlled break
junction (MCBJ) setup in cryogenic vacuum at 4.2 K (see
ref.~\cite{agrait03} for a detailed description). Break junctions
for the metals Au, Ag, Pt and Ni have been investigated with the
molecules H$_{2}$, D$_{2}$, O$_{2}$, C$_{2}$H$_{2}$, CO, H$_{2}$O
and benzene. In most of these cases regular vibration mode spectra
displaying a step down in conductance have been observed, but for
all systems anomalous spectral features as displayed in
Fig.~\ref{fig1} were also found.

In order to admit these molecules to the metal atomic contacts at
4.2 K, the insert is equipped with a capillary that has a heating
wire running all along its interior to prevent premature
condensation of the gasses. The amount of gas admitted is of order
10 $\mu$mol. Previous measurements have clearly demonstrated that
it is possible to capture a single molecule in the atomic junction
and measure its vibration modes \cite{smit02,djukic05}.

\begin{figure} [t!]
\includegraphics[width=8.8cm, angle=0] {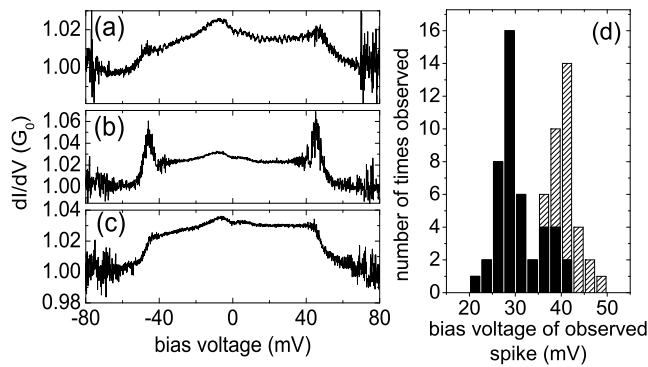}
\caption{\label{fig2} (a) to (c) display the gradual change of
dI/dV spectra of a Pt-D-D-Pt contact; (b) and (c) show spectra
taken for the same junction as (a) after stretching  by 0.1 \AA~
and 0.2 \AA, respectively. Histogram (d) of peak positions in
spectra of junctions for Au-H$_{2}$ (hatched bars) and Au-D$_{2}$
(filled bars). All data were obtained at T = 4.2K.}
\end{figure}

Figure \ref{fig1}(a) shows a dI/dV spectrum of a single CO
molecule contacted by Pt leads. The observation of spikes that
dominate the spectrum is surprising. Pt-CO-Pt contacts at a
conductance of $G\simeq 1 G_{0}$, as for Pt-H-H-Pt contacts
\cite{smit02}, typically display step-like decreases in dI/dV when
conduction electrons are backscattered due to energy loss in
exciting a local vibration mode \cite{djukic06}.  Figure
\ref{fig2}(c) shows an example of a regular dI/dV spectrum for a
Pt-D-D-Pt contact. The I-V curve of Fig.~\ref{fig1}(b) shows clear
fluctuations in the voltage window in which the peaks in dI/dV
appear. At high voltages the fluctuations have disappeared and the
junction displays a slightly higher conductance.

We have evidence that the anomalies are related to vibration modes
of the single molecule. Firstly, for the Pt-H$_2$ system the
spikes in dI/dV typically appear at the same energies as the
transverse vibration mode energies of a hydrogen molecule
\cite{djukic05}. Figures \ref{fig2}(a) to (c) show a sequence of
spectra taken when a Pt-D-D-Pt contact was slightly stretched over
0.2 \AA. The small bumps on the vibration mode shoulders in
Fig.~\ref{fig2}(a) at $\pm$ 48 mV evolve into full peaks in
Fig.~\ref{fig2}(b) and subsequently into a regular vibration mode
spectrum for a deuterium molecule in between Pt contacts in
Fig.~\ref{fig2}(c). Additionally we have introduced hydrogen to
atomic gold contacts and subsequently measured dI/dV. The spectra
show similar anomalous spikes as for Pt-H$_2$. The positions of
the spikes observed in dI/dV spectra for Au-H$_2$ and Au-D$_2$ are
shown as histograms in Fig.~\ref{fig2}(d). It can be clearly seen
that the histogram maximum for Au-D$_2$ at 29meV, has shifted by
the square root of the mass ratio from 42meV for Au-H$_2$. From
the evidence in Fig.~\ref{fig2} it can be concluded that the
observed singularities are associated with the molecular vibration
mode energies in the contact.

Spikes in dI/dV point to abrupt changes in the conductance of the
contacts in small voltage windows. Since similar spikes have been
observed for different molecules this suggests that it involves a
general feature of single molecule junctions. The `noise' visible
in Fig.~\ref{fig1}(b) is a result of two-level fluctuations, that
are excited at high bias. However, they are apparently not excited
directly in view of the connection with vibration mode energies
shown in Fig.~\ref{fig2}. Reference \cite{gaudioso00} has shown
that by exciting internal vibrations of a pyrrolidine molecule on
a metal surface, the molecule can flip to a different
conformation. The different conformations have different
conductances leading to a dip in dI/dV. This forms a basis for our
proposed explanation.

We attribute the abrupt changes in contact conductance to abrupt
switching between two slightly different local geometrical
configurations that have different conductances. Current induced
two-level systems in mesoscopic contacts have been studied in
detail in the past (see
e.g.~\cite{keijsers95,ralph95,halbritter04}). In those cases
spikes are only observed in the second derivative
d$^{2}$I/dV$^{2}$, in contrast to spikes in dI/dV in our
measurements. Since the anomalous spikes appear around the
vibration mode energies, we assume that the switching is induced
by intra-molecular vibrations. STM studies on single molecules on
metal surfaces have clearly shown, e.g., that by exciting an
internal vibration mode of a molecule the lateral hopping rate of
a CO molecule \cite{komeda02}, the rotation of a O$_{2}$ molecule
\cite{stipe981} or of a C$_{2}$H$_{2}$ molecule \cite{stipe982}
can be increased by two orders of magnitude. In our case the
molecules are chemically linked by two leads preventing the
molecules of jumping away and the current is much higher, leading
to orders of magnitude higher rates.
\begin{figure} [b!]
\includegraphics[width=8cm,angle=0] {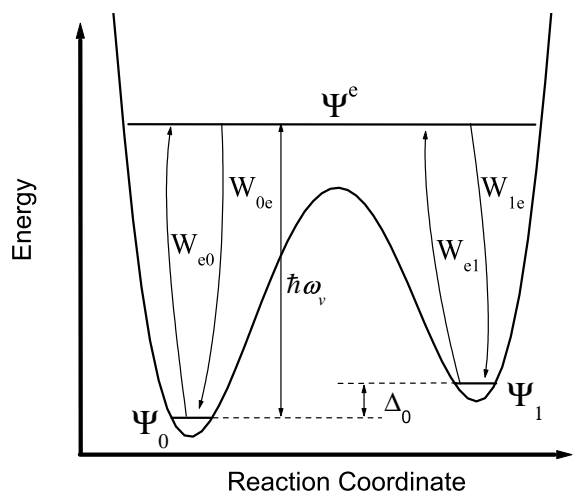}
\caption{\label{fig3} Energy landscape of a molecule in an atomic
contact. The molecule can be vibrationally excited and relax into
the other energy minimum.}
\end{figure}

We derive a simple model of a vibrationally induced two-level
system (VITLS) that reproduces the shape of the anomalous spikes
in our dI/dV spectra very well. The model is schematically
depicted in Fig.~\ref{fig3}. The molecule experiences a double
well potential landscape with ground states $\Psi_{0}$ and
$\Psi_{1}$ in the two energy minima. The two energy minima are
separated by a large potential barrier that prevents tunneling
between the two ground states. The molecule in the contact can be
vibrationally excited by conduction electrons to the state
$\Psi^{e}$. In order to displace the molecule, the barrier for the
reaction coordinate has to be overcome. The direction of the
internal vibration needs to be in the direction of the reaction
coordinate (i.e the movement) of the molecule \cite{persson02}. In
the case of the Pt-D$_2$ junctions the spikes have only been
observed at energies corresponding to vibrations perpendicular to
the current direction (i.e. transverse modes). The longitudinal
vibration mode at about 90 meV \cite{djukic05} never showed
anomalous spikes. In our quasi one-dimensional junctions the
transverse modes are the ones that have the proper reaction
coordinates to induce changes in the geometry of the junctions.

The expression for the current passing through the junction can be
approximately written as:
\begin{eqnarray}\label{current}
I(V) = \left( n_{0} \sigma_{0} + n_{1}\sigma_{1} +
n_{e}\frac{(\sigma_{0} + \sigma_{1})}{2}\right) V
\end{eqnarray}
We assume that the conductance $\sigma_{0}$ of well 0 and
$\sigma_{1}$ of well 1 are different. A step in $I(V)$ and thus a
peak in dI/dV can be obtained once $\Psi_{1}$ becomes populated.
The total current is determined by the sum of the products of the
occupation numbers with their respective conductances. For
simplicity we take the average of $\sigma_{0}$ and $\sigma_{1}$ to
be the conductance of the excited state. In order to determine the
shape of the dI/dV spectrum, the dependence of $n_{0}$, $n_{1}$
and $n_{e}$ on $V$ has to be determined. Let us therefore consider
dynamic equilibrium of the occupation numbers of all levels
involved in the system as displayed in Fig.~\ref{fig3} in terms of
the transition rates,
\begin{eqnarray}\label{rates}
\begin{array}{ll}
\frac{dn_{0}}{dt} = -W_{e0} + W_{0e} = 0 \\
\\
\frac{dn_{1}}{dt} = -W_{e1} + W_{1e} = 0 \\
\\
\frac{dn_{e}}{dt} = -W_{0e} - W_{1e} + W_{e0} + W_{e1} = 0  \\
\\
n_{0} + n_{1} + n_{e} = 1 .
\\
\end{array}
\end{eqnarray}
Here $W_{fi}$ represent the transition rates from states $i$ to
states $f$. The energy difference between the two ground states
(Fig.~\ref{fig3}) we take to be $\Delta_{0}$. The vibration mode
energy $\hbar \omega_{v}$ is much larger than $\Delta_{0}$ and
$k_{B}T \leq \Delta_{0}$ to keep the system well defined in the
ground state for low bias voltages. The transition rates $W$ can
now be determined. Consider as an example the rate $W_{e0}$:
\begin{eqnarray}\label{rates2}
\begin{array}{ll}
 W_{e0} = n_{0} \int \limits_{-\infty}^{+\infty}
\gamma f(E,eV) \left[1-f(E - \hbar\omega_{v},eV) \right]dE  \\
\end{array}
\end{eqnarray}
We assumed here that the electron density of states is flat over
the energy window under consideration. Where $\gamma$ is the
electron-phonon coupling and $f(E,eV)$ are the non-equilibrium
electron distribution functions. All other rates can be expressed
in a similar way. After considering the low temperature limit,
$\frac{\hbar\omega_{v}}{k_{B}T}$,
$\frac{(\hbar\omega_{v}-\Delta_{0})}{k_{B}T}>>1$ and limiting to
positive biases only we can solve the integrals of
Eq.~\ref{rates2}:
\begin{eqnarray}\label{integrals}
\int \limits_{-\infty}^{+\infty}f(E,eV) \left[1-f(E+\Delta
E,eV)\right]dE = \nonumber \\
  \Delta E + \frac{1}{4}\frac{\Delta E -eV}{e^{(\Delta
   E-eV})/k_{B}T -1},  \quad \mbox{when $\Delta E >0,$} \\
  \frac{1}{4}\frac{-\Delta E-eV}{e^{(-\Delta E-eV)/k_{B}T}-1},
   \quad \mbox{ when $\Delta E <0$} \nonumber
\end{eqnarray}
With the results of Eq.~(\ref{integrals}) we can calculate the
rates of Eq.~(\ref{rates}). The set of linear equations can be
solved analytically and the resulting occupation numbers can be
inserted in the expression for the current~(\ref{current}).
\begin{figure} [t!]
\includegraphics[width=9cm,angle=0] {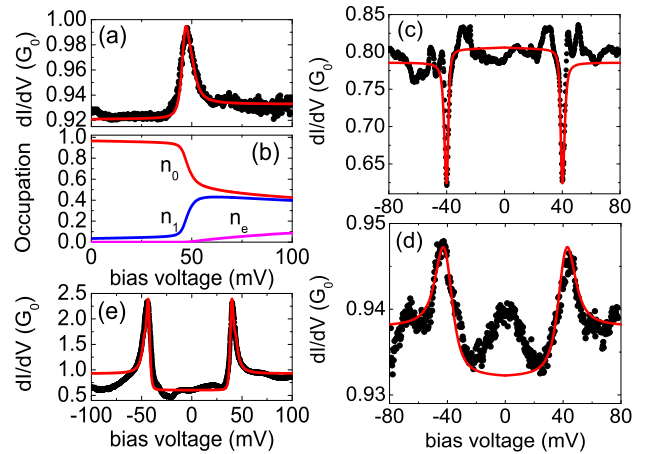}
\caption{\label{fig4} (Color online) Several fits of VITLS for
single molecule junction contacts and atomic chains: (a) H$_{2}$
in a Pt contact; (b) Evolution of occupation numbers n$_{0}$
(red), n$_{1}$ (blue), n$_{e}$ (magenta) for the fit of (a); (c)
H$_{2}$ in a Au atomic chain; (d) H$_{2}$ in Au atomic chain; (e)
O$_{2}$ in a Ni contact. The experimental temperature was 5 K in
all cases except in (d) where the contact was heated to 20 K. For
fit parameters see~\cite{fits}.}
\end{figure}
Figure~\ref{fig4}(a) shows a measurement and fit for a H$_{2}$
molecule in a Pt contact, while Fig.~\ref{fig4}(b) shows the
evolution of the occupation numbers, which clearly shows a sharp
drop in n$_{0}$ and rise in n$_{1}$ once the vibration mode energy
is reached. Figure~\ref{fig4}(c) and (d) show spectra and fits for
H$_{2}$ in a Au atomic chain at different temperatures. In
Fig.~\ref{fig4}(d) the regular phonon mode of the gold atoms in
the chain at $\pm 15$meV can be clearly resolved next to the VITLS
spikes. Figure~\ref{fig4}(c) shows clear negative spikes even when
large conductance fluctuations are present. These conductance
fluctuations arise from interference of electron waves that
scatter on defects and impurities close to the contacts
\cite{ludoph00}. Often, the regular vibration mode signal is
masked by these fluctuations for conductances $<$ 1 G$_{0}$, but
the spikes clearly show. Figure~\ref{fig4}(e) shows a spectrum and
fit for O$_{2}$ in a Ni contact, and a fit was also added in
Fig.~\ref{fig1}(a), indicating that the model is not limited to
hydrogen. The energy splitting between the two ground states
$\Psi_{0}$ and $\Psi_{1}$ is 1--3meV and is clearly larger than
the fitted temperatures, which are about 0.3--0.6meV ($\sim
4$--7K) in all cases except for the spectrum of
Fig.~\ref{fig4}(d), for which the fitted temperature is 1.6meV
($\sim$20K) in agreement with the bath temperature. This implies
that for V$\ll$V$_{m}$ the system is almost entirely in state
$\Psi_{0}$ (see Fig.~\ref{fig4}(b)), which is a prerequisite for
obtaining a sufficiently large decrease of n$_{0}$ and increase of
n$_{1}$ when V reaches V$_{m}$.

The model described above involves vibrational excitation above
the barrier. In the case of hydrogen it is also possible that the
molecule tunnels through the reduced effective barrier after
excitation. We have elaborated a similar model for the case
involving a weak tunnel coupling between excited states in the
potential wells, which leads to the emergence of two
superpositions of the excited states \cite{darkothesis}. The
resulting fits are not qualitatively different. Furthermore, we
have investigated the possibility of double excitations in the
potential wells within the framework of our model. The resulting
fits could not reproduce the experimental data, due to a much
slower population of the states of well 1.

As illustrated by Fig.~\ref{fig2} the conditions for obtaining a
measurable VITLS signal can be obtained by control of the MCBJ: It
is possible by pulling and pushing the contact very gently to
explore the potential landscape for the molecule so that the
proper conditions for a VITLS can be obtained.

Recently STM measurements on a hydrogen covered Cu(111) surface
yielded very similar anomalous dI/dV spikes \cite{gupta05}. The
mechanism is probably related but not identical to ours because
the peak positions showed strong dependence on hydrogen coverage
of the surface. Moreover, the observed two-level fluctuations were
observed to remain in the $\sigma_{1}$-state above the threshold
voltage. Possibly the geometry of the experiment allows for the
molecules to be excited into positions away from the STM tip,
after which they are slow to return.

The IETS measurements by Wang {\it et al.}~\cite{wang04} on a
self-assembled monolayer of octanedithiols have been discussed
\cite{galperin04} in view of the unexpected shape of the features
in d$^2$I/dV$^2$. Several of the stronger features have the shape
of peaks in the first derivative, dI/dV, but still at the energies
of known vibration mode energies for the molecules. Our model
produces a good fit to the data by Wang {\it et al.} but the
temperature in the fit is much higher than the experimental
temperature. This may be attributed to inhomogeneous broadening in
the ensemble of molecules measured.

We conclude that the observed spikes in dI/dV measurements on
single molecule junctions are most probably due to vibrationally
induced two-level systems. Our simple model fits the data very
well. It will be interesting to investigate potential energy
landscapes for single molecule junctions in atomistic models in
order to obtain a more concrete picture of the states involved.
The VITLS acts as an intrinsic amplification mechanism providing a
strong spectroscopic signature for local vibration modes and the
anomalous spikes in dI/dV are expected to be a general feature of
single molecule junctions. It can therefore be exploited as a new
spectroscopic tool in analogy to the "action spectrum" proposed by
Komeda {\it et. al.}~\cite{komeda02}. Specifically, Au atomic
chains with H$_2$, that very rarely show regular vibration mode
spectra, have a vibration mode around 42 meV, as seen from
Fig.~\ref{fig2}(d).

We would like to thank W.Y. Wang, M.A. Reed, J.A. Gupta and A.J.
Heinrich for valuable discussions and communicating their dI/dV
spectra. This work is part of the research program of the
``Stichting FOM,'', partially financed through the SONS programme
of the European Science Foundation, which is also funded by the
European Commission, Sixth Framework Programme, and was also
supported by the European Commission TMR Network program DIENOW.

\end{document}